# Spontaneous imbibition in porous media: from pore scale to Darcy scale

Chao-Zhong Qin[1,2,*], Xin Wang[1,2], Mahmoud Hefny[3,4], Jianlin Zhao[5], and Bo Guo[6]

1. State Key Laboratory of Coal Mine Disaster Dynamics and Control, Chongqing University, Chongqing, China
2. School of Resources and Safety Engineering, Chongqing University, Chongqing, China
3. Geothermal Energy and Geofluids, Institute of Geophysics, ETH Zürich, Switzerland
4. Geology Department, South Valley University, Egypt.
5. Department of Mechanical and Process Engineering, ETH Zürich, Switzerland
6. Department of Hydrology and Atmospheric Sciences, University of Arizona, Tucson, AZ, USA

**Abstract**

Spontaneous imbibition has been receiving much attention due to its significance in many subsurface and industrial applications. Unveiling pore-scale wetting dynamics, and particularly its upscaling to the Darcy scale are still unresolved. In this work, we conduct image-based pore-network modeling of cocurrent spontaneous imbibition and the corresponding quasi-static imbibition, in homogeneous sintered glass beads as well as heterogeneous Estaillades. A wide range of viscosity ratios and wettability conditions are taken into account. Based on our pore-scale results, we show the influence of pore-scale heterogeneity on imbibition dynamics and nonwetting entrapment. We elucidate different pore-filling mechanisms in imbibition, which helps us understand wetting dynamics. Most importantly, we develop a non-equilibrium model for relative permeability of the wetting phase, which adequately incorporates wetting dynamics. This is crucial to the final goal of developing a two-phase imbibition model with measurable material properties such as capillary pressure and relative permeability. Finally, we propose some future work on both numerical and experimental verifications of the developed non-equilibrium permeability model.

1. **Introduction**

**I**n porous media research, spontaneous imbibition is capillary-driven invasion of the wetting phase displacing the nonwetting phase, which is governed by the interplay of capillary force and viscous force. As a typical two-phase flow process, it plays an important role in numerous practical problems such as oil production from fractured reservoirs (Kyte, 1962; Morrow & Mason, 2001), remediation of non-aqueous phase liquids in soils (Singh & Niven, 2014), residual trapping in geological carbon dioxide storage (Scanziani et al., 2020), inkjet printing (Wijshoff, 2018), and paper sensors (Rath et al., 2018). Furthermore, spontaneous imbibition has been used to infer wettability of many geological materials (Peng & Xiao, 2017). There have been many fundamental studies of spontaneous imbibition mainly focusing on the predictions of imbibition rate (Akin et al., 2000; Hall & Pugsley, 2020; Kuijpers et al., 2017; Standnes & Andersen, 2017), entrapment of nonwetting phase (Meng et al., 2015), broadening/roughening of wetting front (Gruener et al., 2012; Sadjadi & Rieger, 2013; Soriano et al., 2005; Zahasky & Benson, 2019), and Darcy-scale capillary pressure and relative permeability (Alyafei & Blunt, 2018; Haugen et al., 2014; Mason & Morrow, 2013).

Spontaneous imbibition can be categorized into cocurrent spontaneous imbibition and countercurrent spontaneous imbibition, according to the way the wetting phase enters and leaves a medium. In cocurrent spontaneous imbibition, the wetting phase enters a porous medium through one boundary, while the nonwetting phase is displaced from the medium via other boundaries. By contrast, in countercurrent spontaneous imbibition, the wetting and nonwetting phases respectively enter and leave a porous medium via the same boundary. In practice, they may coexist and switch from one to the other depending on the viscosity ratio and the domain size (Haugen et al., 2014). Overall, the two imbibition processes are distinct in pore-filling mechanisms, imbibition dynamics, and Darcy-scale material properties. For countercurrent spontaneous imbibition (e.g., the scenario of OEO), experiments show that the imbibition rate commonly is linearly proportional to the square root of imbibition time, and many scaling groups have been developed (Mason & Morrow, 2013). However, it is still unclear how capillary backup pressure and backup saturation evolve in time, which are important input parameters to the Darcy-scale modeling. Countercurrent spontaneous imbibition may be close to quasi-static, however, its difference from quasi-static imbibition is still unclear so that it is not confident to use conventionally measured capillary pressure and relative permeability curves in the Darcy-scale modeling.

There is a long history of research on cocurrent imbibition, pioneered by Lucas and Washburn's work on imbibition in a capillary tube (Washburn, 1921). By treating a porous medium as a bundle of capillary tubes, various analytical/semi-analytical models have been developed to predict the imbibition rate (Cai et al., 2021; Standnes, 2010). Likewise, by assuming a sharp wetting front and neglecting the nonwetting-phase resistance and gravity, the wetting penetration depth can be simply given by the single-phase Darcy model as $L = \sqrt{2kp^c/\varepsilon\mu}\sqrt{t}$, where ε is the porosity, μ is the dynamic viscosity, $k$ is the wetting permeability depending on the entrapment of the nonwetting phase, $p^c$ is the Darcy-scale capillary pressure, and $t$ is the imbibition time. To quantitatively predict the imbibition rate, one needs to know wetting permeability and capillary pressure in advance. However, it is still unclear how to relate them to conventional material properties of a porous medium such as pore-size distribution, wettability, quasi-static capillary pressure curve, and relative permeability measured under steady state (Akbarabadi & Piri, 2013; Gao et al., 2020; Krause & Benson, 2015; Ruprecht et al., 2014). Moreover, in the case of late stage of imbibition or imbibition with low viscosity ratios (e.g., water imbibes into an oil-saturated medium), the assumption of sharp wetting front will not hold any more, so that the two-phase Darcy model needs to be employed (F. Zarandi & Pillai, 2018).

There have been many studies of the Darcy-scale modeling of cocurrent spontaneous imbibition, primarily for the air-water system in which water is the wetting phase. It is acknowledged that dynamics of self-determined spontaneous imbibition causes Darcy-scale capillary pressure and relative permeability to deviate from their values at equilibrium. Therefore, in most studies, either assumed or experimentally fitted relationships different from the ones measured under steady-state/quasi-static conditions were usually used (Alyafei et al., 2016; Schmid et al., 2016; Suo et al., 2019; Zahasky & Benson, 2019). It is unclear how they are related to the quasi-static counterparts and also pore-scale events. Furthermore, those fitted relationships cannot be extended to other fluid systems of imbibition. Regarding the effect of dynamics (or non-equilibrium) on capillary pressure and relative permeability, there are a few pioneering studies including the well-known tau term in capillary pressure proposed by Hassanizadeh and Gray (Hassanizadeh & Gray, 1990; Zhuang et al., 2017), and the effective saturation proposed by (Barenblatt et al., 2003). The benefit of those models is that they may physically link the non-equilibrium effect to conventional capillary pressure and relative permeability. However, relevant studies in spontaneous imbibition are quite rare.

There are some other challenges in the prediction of spontaneous imbibition, such as the effects of pore heterogeneity, initial wetting saturation, gravity, and mixed-wettability on the imbibition rate and entrapment of the nonwetting phase (or residual saturation) (Bartels et al., 2019). In this regard, pore-scale studies will be indispensable. Given the fact that pore-scale events in spontaneous imbibition, particularly in cocurrent spontaneous imbibition, are so fast that current imaging techniques may not capture them, the pore-scale modeling will be extremely useful. Previous pore-scale modeling of spontaneous imbibition have been limited to synthetic 2D porous media or a small porous domain much smaller than the REV size (Liu et al., 2020; Rokhforouz & Akhlaghi Amiri, 2018). As a result, the obtained pore-scale information cannot be upscaled to Darcy-scale capillary pressure and relative permeability. Alternatively, the efficient pore-network modeling is promising (Aghaei & Piri, 2015; Chen et al., 2020; Joekar-Niasar et al., 2010). In this work, we use our in-house pore-network simulator, which has been verified to direct numerical simulations (Qin et al., in preparation), to study cocurrent spontaneous imbibition and quasi-static imbibition in both homogenous and heterogeneous porous media. Based on the pore-scale results, as a first attempt, we aim (1) to illustrate the effect of pore-scale heterogeneity on imbibition dynamics and nonwetting entrapment, (2) to elucidate the difference between cocurrent spontaneous imbibition and quasi-static one in terms of pore-filling mechanisms and nonwetting entrapment, and (3) last but not least to develop a relative permeability model for the wetting phase in spontaneous imbibition, which takes into account imbibition dynamics.

The remainder of the paper is organized as follows. In section 2, we will introduce the used μCT data of sintered glass beads and Estaillades carbonate. The used pore-network model and case studies will be presented in Section 3, which is followed by results and discussion in Section 4. Finally, we will draw main conclusions and an outlook in Section 5.

## 2. Porous samples: sintered glass beads and Estaillades carbonate

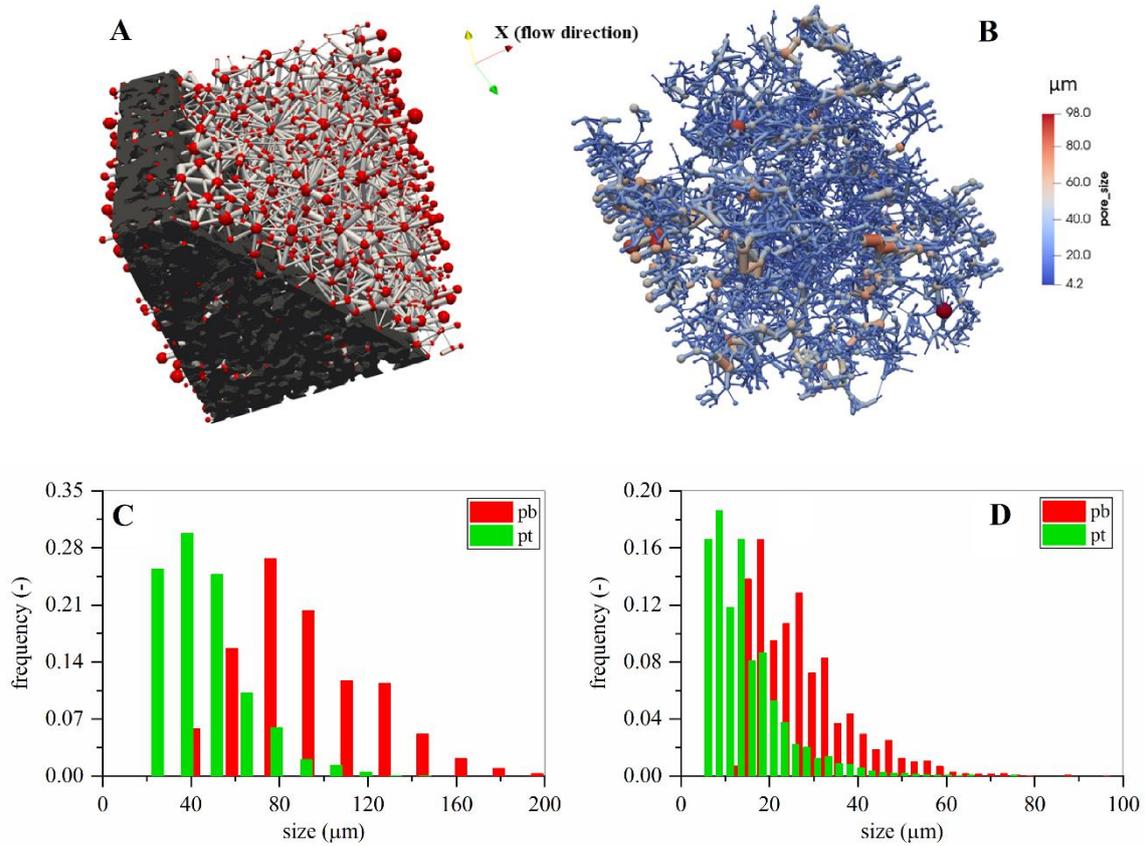

**Fig. 1:** Details of the two pore networks: (A) the extracted pore network and solid phase of the pore medium of sintered glass beads; (B) the extracted pore network of Estaillades, the colormap shows the distribution of inscribed radii; (C) the pore-body inscribed radius and the pore-throat inscribed radius distributions of sintered glass beads; and (D) the pore-body inscribed radius and the pore-throat inscribed radius distributions of Estaillades.

We use sintered glass beads and Estaillades carbonate (Muljadi et al., 2016) as our porous samples. PoreSpy is used to extract their pore networks (Gostick, 2017). As shown in Fig. 1A, the porous medium of sintered glass beads features homogeneous, which also resembles highly-permeable sandstones in terms of complex pore structures. Fig. 1B shows the extracted pore network of Estaillades carbonate, which features strong pore-scale heterogeneity. Meanwhile, it is found isolated pore spaces account for around 14% of the overall pore volume. The pore-size distributions of the two samples are seen in Fig. 1C and Fig. 1D, in which the distribution of sintered glass beads is closer to lognormal.

The details of the pore networks and geometrical parameters are given in Table 1. We run Lattice-Boltzmann simulations of each pair of watersheds in the pore networks to obtain its conductance, and assign it to the corresponding pore throat (Zhao et al., 2020). A similar practice was also used by (Raeini et al., 2017) in their algorithm of pore-network extraction. The homogenous medium of sintered glass beads is highly permeable with a permeability of 113.2 D, while the heterogeneous Estaillades has a much smaller permeability of 0.303 D. Half of the Estaillades was used by (Muljadi et al., 2016) to study the heterogeneity on non-Darcy flow behavior, in which a permeability of 0.172 D was reported.

Given a mean pore size (i.e. radius) and permeability, the tortuosity may be calculated by the empirical model: $k = \varepsilon r^2 / 8\tau^2$ (Kuijpers et al., 2017). In such way, the tortuosity values of sintered glass beads and Estaillades are 1.8 and 6.9, respectively. Overall, a wide pore-size distribution combined with poor connectivity (see Fig. S1 for the distributions of coordination number in the supporting information) result in strong pore-scale heterogeneity of Estaillades with high tortuosity.

**Table 1:** Geometrical and physical parameters in the case studies.

| Geometrical parameters | Sintered glass beads | Estaillades |
| --- | --- | --- |
| Voxels | 260×260×260 | 1000×1000×1000 |
| Resolution | 25 μm | 3.31 μm |
| Domain length | 6.5 mm | 3.31 mm |
| Number of pore bodies / pore throats | 2019 / 5225 | 4208 / 7710 |
| Number of inlet / outlet pores | 140 / 137 | 113 / 84 |
| Mean coordination[1] | 5.8 | 3.7 |
| Porosity[2] | 27.8% | 11.3% |
| **Mean pore size[3]** | ***102.4 μm*** | ***32 μm*** |
| Permeability | 113.2 D | 0.303 D |
| Water/air viscosity | $1.0 \times 10^{-3} / 1.79 \times 10^{-5}$ Pa s | |
| Surface tension | 0.073 N/m | |
| Inlet/outlet mixture phase pressure | 0.0/0.0 Pa | |
| Pre-wetting film capillary pressure | $10^6$ Pa | |

1. Excluding inlet and outlet pore bodies and pore throats.
2. Excluding isolated pores, Estaillades has a porosity of 11.3% (close to the value of 10.8% reported in (Muljadi et al., 2016). Estaillades with isolated pores has a porosity of 13.2%.
3. The mean pore size is volume-averaged.

## 3. Pore-network model and case studies

The present dynamic pore-network model has been developed for image-based modeling of spontaneous imbibition (Qin et al., in preparation). Multiform idealized pore elements are used to represent complex pore structures. Distinct from previous dynamic pore-network models of two-phase flow in porous media, our model incorporates a local rule of the competition of Main Terminal Meniscus (MTM) move and arc meniscus (AM) filling. This local rule, to a certain extent, describes the behavior of two-phase interfaces in connected pore spaces in imbibition. Moreover, we calibrate the single-phase conductance of each pair of connected pores by LBM simulations, which makes our model potential to be quantitative (Zhao et al., 2020). As many others (Joekar-Niasar et al., 2010; Sinha & Wang, 2007; Thompson, 2002), we solve the following governing equations of two-phase flow in a pore network:

$$V_i \frac{ds_i^\alpha}{dt} = -\sum_{j=1}^{N_i} K_{ij}^\alpha (p_i^\alpha - p_j^\alpha) \qquad \alpha = \{n, w\} \tag{1}$$

$$\sum_{j=1}^{N_i}(K_{ij}^n + K_{ij}^w)(\bar{p}_i - \bar{p}_j) = -\sum_{j=1}^{N_i}\{[K_{ij}^n s_i^w - K_{ij}^w(1-s_i^w)]p_i^c + [K_{ij}^w(1-s_j^w) - K_{ij}^n s_j^w]p_j^c\} \tag{2}$$

where $i$ is the pore-body index, $ij$ is the pore-throat index, $n$ and $w$ indicate the nonwetting and wetting phases, respectively, $N_i$ is the coordination number of pore body $i$, $V$ [m$^3$] is the volume, $s$ [-] is the saturation, $K$ [m$^3$/Pa/s] is the conductivity, and $\bar{p}$ [Pa] is the mixture pressure defined $\bar{p} = p^n s^n + p^w s^w$. The gravity is neglected. The capillary pressure in pore body $i$ is defined as $p_i^c = p_i^n - p_i^w$.

The governing equations (1, 2) are numerically solved for the primary variables: wetting saturation, $s_i^w$, and mixture pressure, $\bar{p}_i$. At the end of each time step, capillary pressure and phase conductivity are updated based on the primary variables. The details of the calculation of constitutive relations of the used idealized pore elements are presented elsewhere (Qin & van Brummelen, 2019).

For the modeling of quasi-static imbibition, we use our in-house quasi-static pore-network model which considers key two-phase displacement mechanisms such as cooperative filling of pore bodies, piston-type move, and snap-off (Al-Futaisi & Patzek, 2003; Hefny et al., 2020). The model has been developed following (Patzek, 2001; Valvatne & Blunt, 2004), which has the potential to be quantitative. We assign all pore volume to pore bodies so that pore throats are volumeless, since we use the concept of watershed to prepare our pore networks (Gostick, 2017).

To fulfill our objectives, we concentrate on the early stage of cocurrent spontaneous imbibition and the corresponding quasi-static imbibition. We conduct free spontaneous imbibition with one end face in contact with a wetting reservoir and the other end face in contact with a nonwetting reservoir. The surrounding four faces are sealed. Numerically, we depress any countercurrent imbibition at the wetting reservoir. Although the used samples may be mixed-wettability, uniform wettability has been assumed on purpose. A number of case studies under four contact angel values (20°, 40°, 60°, and 80°) and three viscosity ratios (55.9 for the air-water system, 1.0, and 0.1) have been conducted.

## 4. Results and discussion

### 4.1. Effect of pore-scale heterogeneity on spontaneous imbibition

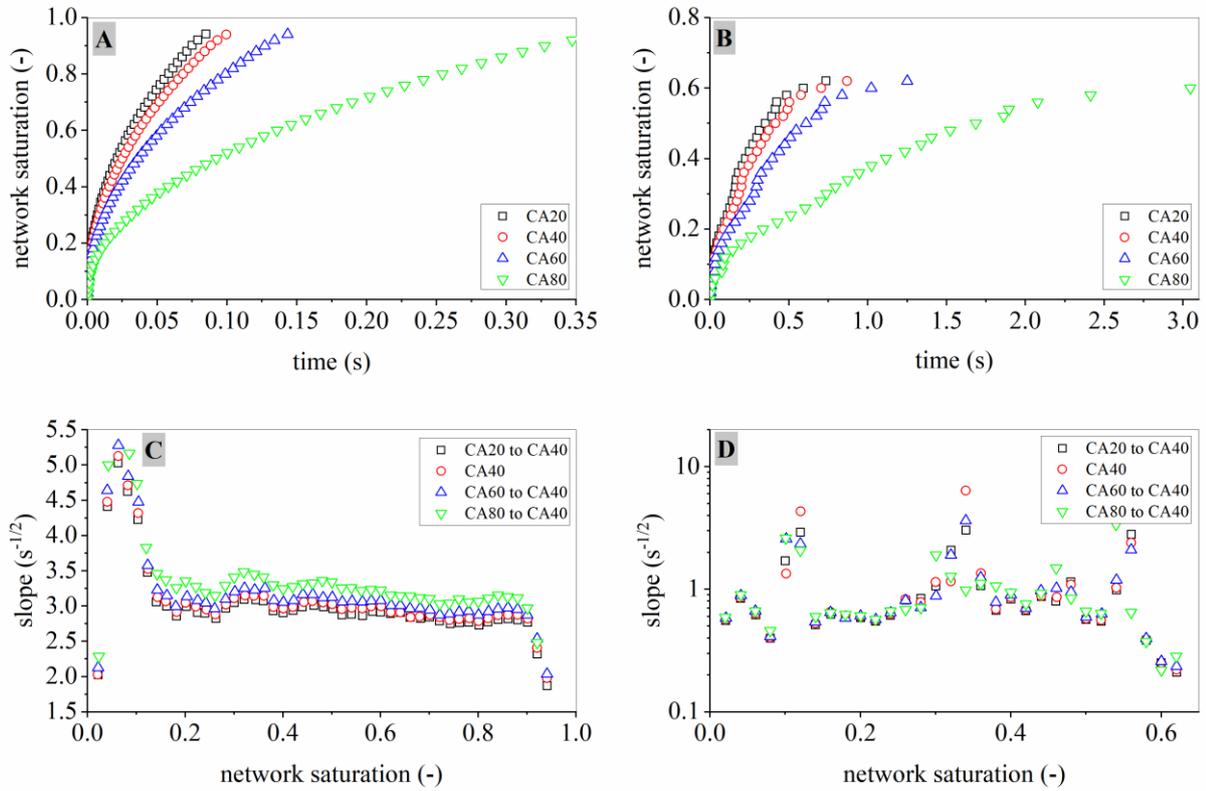

**Fig. 2:** Spontaneous imbibition of water into dry sintered glass beads and Estaillades under different contact angle values: (A) the network water saturation versus the imbibition time for sintered glass beads; (B) the network water saturation versus the imbibition time for Estaillades; (C) the scaled imbibition rate with the contact angle value of 40° versus the network saturation for sintered glass beads; and (D) the scaled imbibition rate with the contact angle value of 40° versus the network saturation for Estaillades. The scaling is done by multiplying the imbibition rate of interest by $\sqrt{\cos 40°/\cos\theta}$ where $\theta$ is the scaled contact angle. The air-water system is considered with the viscosity ratio of 55.9.

Spontaneous imbibition of water into dry sintered glass beads and Estaillades has been conducted under four different contact angle values. Fig. 2A shows the temporal evolutions of the network-scale water saturation in sintered glass beads. The four imbibition curves are smooth due to the pore-scale homogeneity of sintered glass beads. Fig. 2C shows the scaled imbibition rates (i.e., $S/\sqrt{t}$, where $S$ is the network saturation and $t$ is the imbibition time) versus the network saturation, and the reference imbibition rate was obtained under the contact angle value of 40°. It is seen that all imbibition rates more or less lump to a constant except for those at the initial and final stages of imbibition, which indicates the imbibition rate in sintered glass beads is linearly proportional to the square root of imbibition time, and the imbibition rate can be adequately scaled with $\sqrt{\cos\theta}$. Moreover, with respect to the early stage of spontaneous imbibition, the medium size of sintered glass beads approaches its REV size. We observe that residual saturation (i.e. the amount of trapped air) in homogenous sintered glass beads is very low (around 0.06), which is in consistence with experimental data (Kuijpers et al., 2017).

Fig. 2B shows the temporal evolutions of overall water saturation in Estaillades. The pore-scale heterogeneity is clearly reflected in the imbibition curves. A few peaks of the imbibition rate are seen in Fig. 2D. The heterogeneity, however, has little influence on the scaling behavior of the imbibition rate, which indicates the dominance of MTM move in the pore-filling. Compared with sintered glass beads, the REV size of Estaillades should be much larger. Moreover, residual saturation in Estaillades is much higher up to 0.36. This value is in good match with experimental data (one can refer to Fig. 4a in (Alyafei et al., 2016)), which proves the reliability of our model in the prediction of nonwetting entrapment.

Fig. 3 shows pore-scale water distributions in sintered glass beads and Estaillades in the end of spontaneous imbibition. The air-water system was considered with the static contact angle value of 40°. It is seen that a few partially-filled/dry pores spread over the medium in sintered glass beads, while a large amount of dry pores present in Estaillades in the form of individual clusters. Obviously, the pore-scale heterogeneity of Estaillades results in severe entrapment of air. There is no preference for air to be trapped in small pores for both sintered glass beads and Estaillades, which indicates the cofilling of small and large pores at the early stage of imbibition. Moreover, the entrapment of air in heterogeneous Estaillades is primarily due to the poor connectivity of pores, which may be termed as topological trapping. The dynamic imbibition processes are seen in Movies in supplementary support. A sharp wetting front with a few pores roughening moves through sintered glass beads. Cofilling dynamics is clearly seen. In heterogeneous Estaillades, however, the imbibition follows preferential pathways, leaving air in poorly connected regions. Although heterogeneity has pronounced influence on pore-scale imbibition dynamics, it is worth noting that at the REV scale a sharp wetting front holds in both sintered glass beads and Estaillades for the air-water system (Alyafei et al., 2016). The main differences lie in the REV size, residual saturation, roughening of wetting front.

There are two ways to obtain REV-scale capillary pressure. We take the medium of sintered glass beads as an example, given that fact that its REV size is satisfied here. One way is that we calculate interface-averaged capillary pressure directly from the pore-scale results. The other is that we use the Lucas-Washburn-type model, $S = \sqrt{2kp^c/\varepsilon\mu d^2}\sqrt{t}$, where $d$ is the medium length. From the pore-scale results, the imbibition rate, $\sqrt{2kp^c/\varepsilon\mu d^2}$, and the wetting relative permeability, $k =$

$0.87 \times 113.2 \, D$, are known. Together with known material properties, capillary pressure can be calculated. Fig. S3 shows the REV-scale capillary pressure values obtained in the two ways, which are in good match. The capillary pressure at the wetting front has the value of around 700 Pa, which stays at the end of quasi-static main imbibition curve (MIC) shown in Fig. S2.

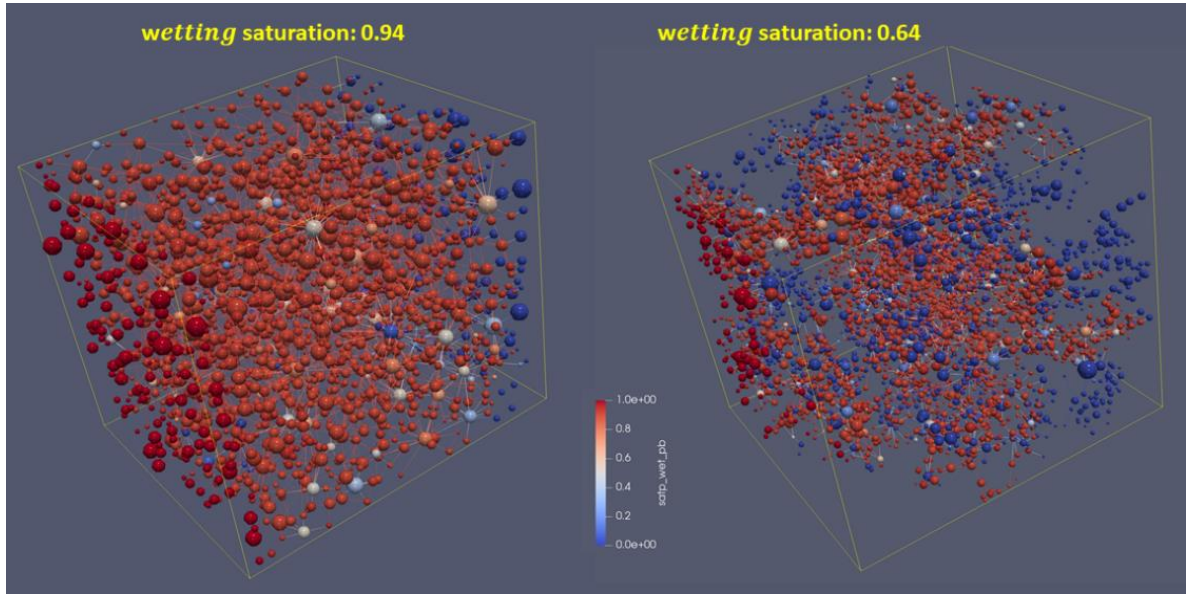

**Fig. 3:** Water distributions in sintered glass beads and Estaillades in the end of spontaneous imbibition. The final overall water saturation values in sintered glass beads and Estaillades are 0.94 and 0.64, respectively. The air-water system was considered with the static contact angle value of 40°.

### 4.2. comparison between quasi-static imbibition and spontaneous imbibition

In the quasi-static modeling, we put the wetting reservoir at the inlet and the nonwetting reservoir at the outlet. We first run the primary drainage to an irreducible saturation of 0.05; then, we run the imbibition to the end (i.e., the nonwetting reservoir pressure reduces to zero). Fig. 4A shows the residual saturation values in sintered glass beads and Estaillades under different contact angle values. Overall, the residual saturation gradually decreases as the contact angle increases, which is much larger than the corresponding one in spontaneous imbibition (see Fig. 2). Furthermore, severer entrapment of the nonwetting phase is seen in Estaillades, e.g., with a residual saturation of 0.86 under the contact angle value of 40°. Fig. 4B, 4C, and 4D show the cooperative-filling events of pore bodies and the snap-off and piston-move events of pore throats, which help us understand the trapping mechanism in quasi-static condition. We observe that in sintered glass beads the number of cooperative-filling events firstly increases, then decreases as the contact angle increases, and an obvious depression of cooperative fillings is seen under the contact angle value of 80°. As expected, the frequency of pore throats with snap-off and piston-move decreases as the contact angle increases shown in Fig. 4D. On one hand, snap-off and piston-move of pore throats benefit cooperative fillings of pore bodies. On the other hand, a high frequency of snap-off and piston-move causes topological trapping of the nonwetting phase, and then reduces the number of cooperative-filling events. Therefore, we may attribute the influence of wettability on the cooperative-filling events to the counteraction of the above two mechanisms. A similar tendency of the cooperative fillings is found in heterogeneous Estaillades. Moreover, we observe that cooperative fillings are tremendously depressed, associated with a sharp reduction of snap-off and piston-move events. Finally, we notice that a high frequency of pore throats with snap-off and piston-move and a low frequency of pore bodies with cooperative fillings (dividing the number of events by the total number of pore bodies) together give rise to severe entrapment of the nonwetting phase in Estaillades.

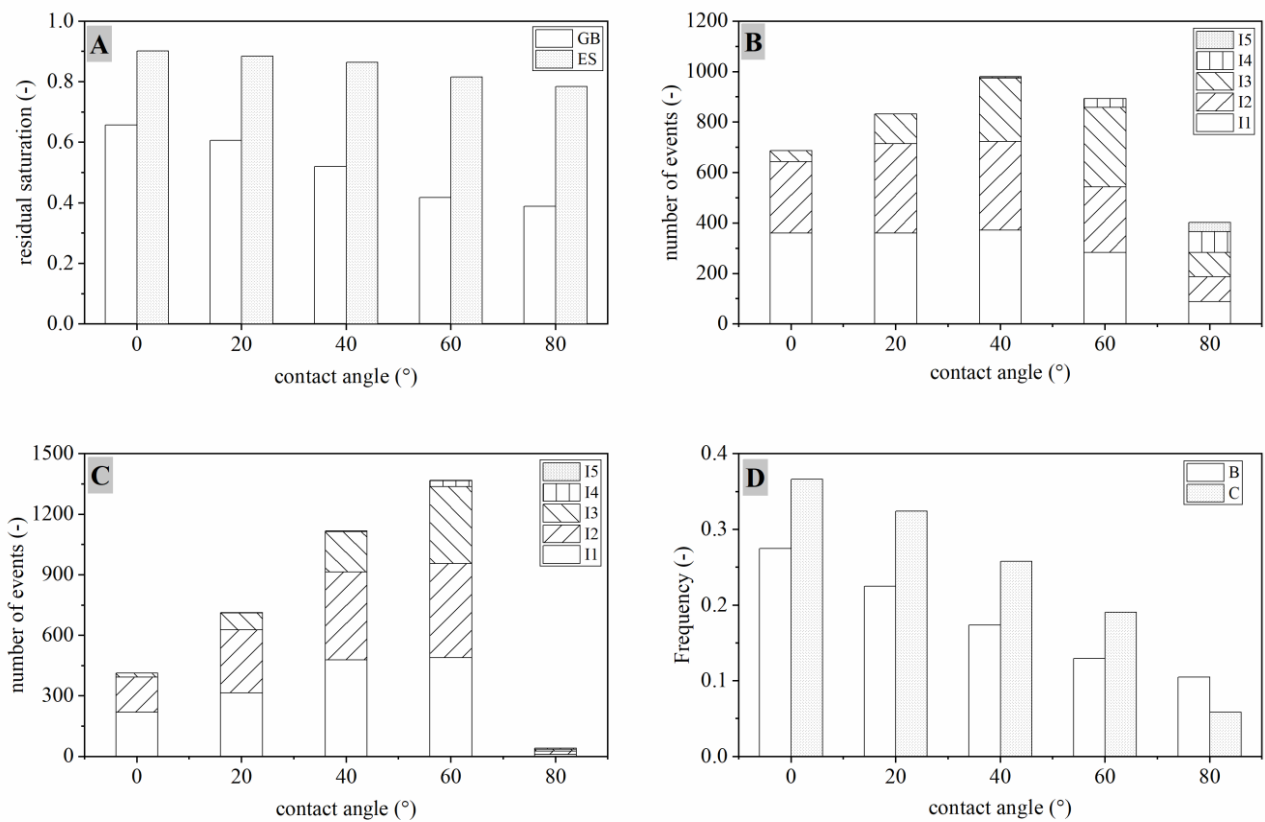

**Fig. 4:** (A) Residual saturation under different contact angle values; (B) the number of cooperative fillings of pore bodies in sintered glass beads, the total number of pore bodies excluding inlet and outlet ones is 1742; (C) the number of cooperative fillings of pore bodies in Estaillades, the total number of pore bodies excluding inlet and outlet ones is 4011; and (D) the frequency of pore throats with snap-off and piston-move. There are five types of cooperative pore-body filling, one can refer to (Patzek, 2001) for the detail.

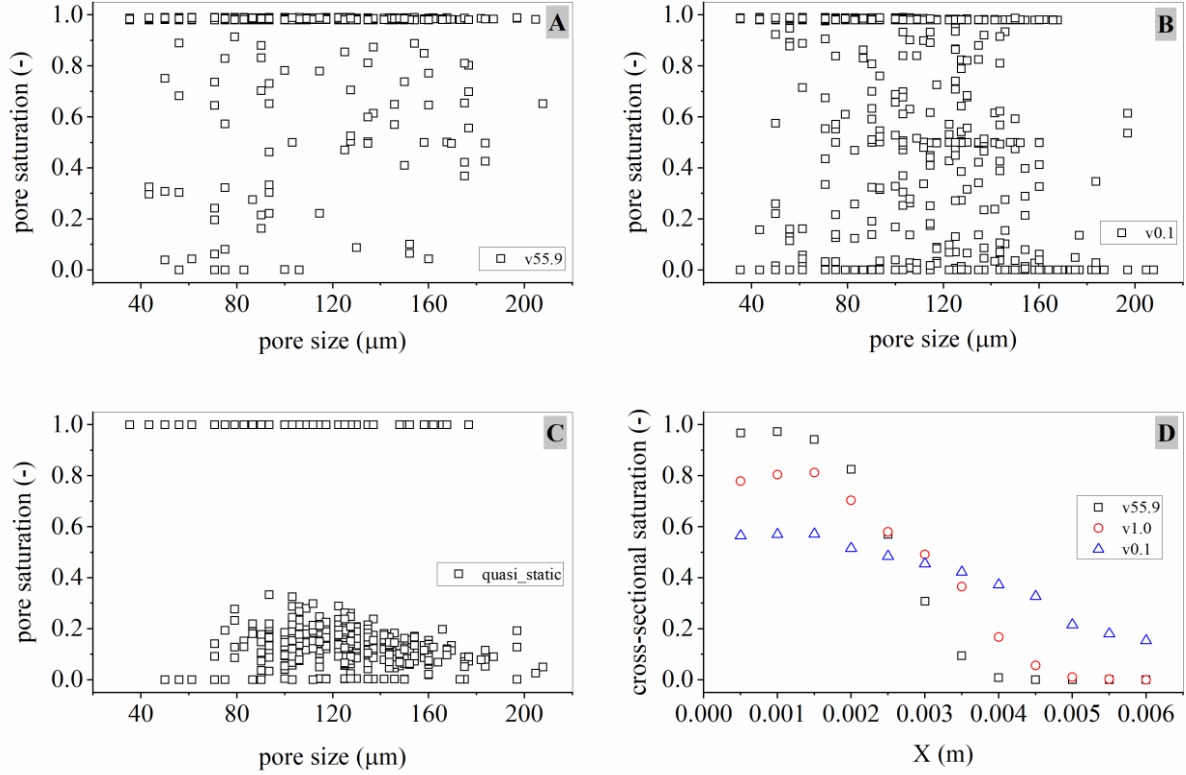

**Fig. 5:** Pore-body saturation values versus pore sizes in sintered glass beads (A) under the viscosity ratio of 55.9 (i.e. the air-water system), (B) under the viscosity ratio of 0.1, and (C) under the quasi-static condition; (D) the cross-sectional saturation profiles along the flow direction at the overall saturation of 0.4 under three viscosity ratios. The static contact angle is 40°.

Pore-filling events depend on imbibition dynamics and the viscosity ratio. According to the pore-size distribution of sintered glass beads (see Fig. 1C), we may categorize the pore bodies into three groups, namely, small pores ranging from 40 µm to 80 µm, intermediate pores from 80 µm to 120 µm, and large pores over 120 µm which account for a relatively small portion. Fig. 5A shows the pore-body saturation values versus pore sizes in the end of spontaneous imbibition for sintered glass beads with $M$ of 55.9. It is seen that only a few partially-filled pores present in the medium, due to the afore-mentioned cofilling mechanism. The wetting-phase resistance assists in the cofilling, while depressing the roughening growth of the wetting front. For a small viscosity ratio of 0.1 in Fig. 5B, we observe much more dry and partially-filled pores across the medium, with the majority of intermediate and large pores. The resultant residual saturation is around 0.38, and the wetting mechanism lies in the interplay of capillary force and viscous resistance of the nonwetting phase. The case of quasi-static imbibition is shown in Fig. 5C. As expected, the wetting filling is predominant in small pores, and it follows the mechanism of percolation-invasion. The resultant residual saturation reaches its peak of around 0.52. For the details of pore-scale wetting distributions in the end of spontaneous imbibition and quasi-static imbibition, one can refer to Fig. S4. Finally, the effect of viscosity ratio on the front roughening is shown in Fig. 5D. Because of different pore-filling mechanisms, reducing viscosity ratio causes the increase of the nonwetting entrapment, and broadens the wetting front. We notice that for a small viscosity ratio, a viscous-fingering pattern, often seen in drainage, is not observed in spontaneous imbibition (see Fig. S4). This is probably because that the propagation of wetting front is not solely dependent viscous resistance of the nonwetting phase.

### 4.3. Relative permeability of the wetting phase

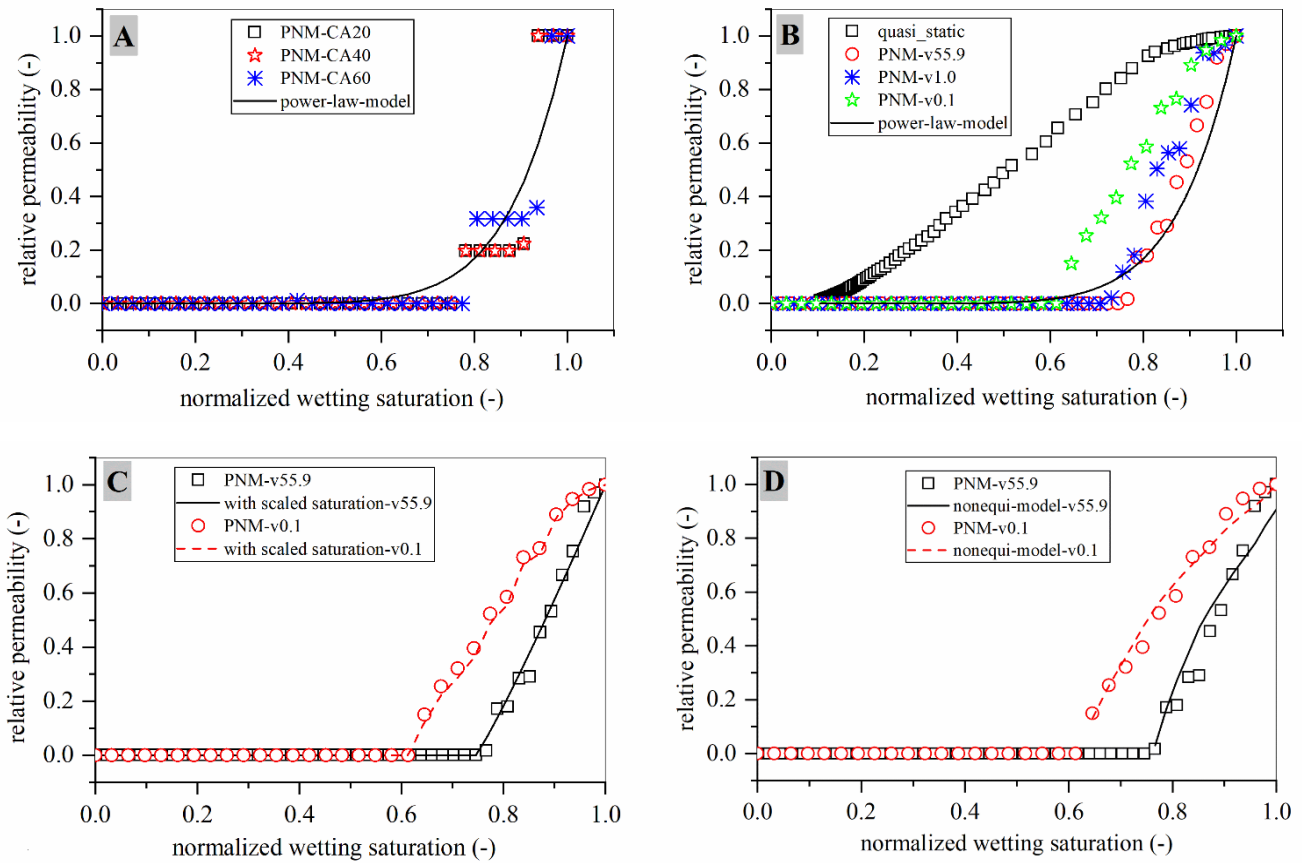

**Fig. 6:** (A) Wetting relative permeability under different contact angle values for Estaillades by the PNM (pore-network model) and the power-law model (the exponent is set to 8), the viscosity ratio is 55.9 for the air-water system; (B) the wetting relative permeability under different viscosity ratios, quasi-static, and by the power-law model (the exponent is set to 8) for sintered glass beads under the contact angle value of 40°; (C) the wetting relative permeability for sintered glass beads under a high and a low viscosity ratios predicted by the PNM and the power-law model with scaled saturation (the exponent is set to 1.1); and (D) the wetting relative permeability for sintered glass beads under a high and a low viscosity ratios predicted by the PNM and the non-equilibrium model (the exponent is set to 1.1).

Two approaches are available to determine relative permeability of a porous medium. One is called the steady-state method, in which a narrow range of wetting saturation can be experimentally achieved (Akbarabadi & Piri, 2013). In the other approach, relative permeability is numerically solved based on the information of pore-scale phase distributions, which may be obtained by the imaging technique (Berg et al., 2016) or the pore-scale modeling (Al-Futaisi & Patzek, 2003). Here, we use the latter approach, and pore-scale phase distributions are from the dynamic and quasi-static pore-network modeling. Fig. 6A shows the water relative permeability under different contact angle values for Estaillades. It is seen that the water relative permeability is not sensitive to the contact angle under the assumption of uniform wettability. Due to the heterogeneity and the limit of the study domain, smooth relative permeability curves are not obtained. Nevertheless, the relative permeability curve by the power-law model with an exponent of 8 locates among those relative permeability values. It is also seen that for the air-water system (i.e., the viscosity ratio is 55.9) the power-law model with an exponent of 8 can moderately fit the water relative permeability for sintered glass beads as shown in Fig. 6B. From the pore-scale perspective, we could understand why many Darcy-scale models of spontaneous imbibition entail a large exponent value in their power-law or Corey expressions for water relative permeability (Alyafei et al., 2016; Schmid et al., 2016; Suo et al., 2019), in order to produce sharp wetting fronts. The influence of imbibition dynamics on wetting relative permeability is pronounced in Fig. 6B. It is seen that imbibition dynamics causes the deviation of the wetting relative permeability away from its quasi-static curve (the relative permeability of primary drainage and main imbibition for sintered glass beads is given in Fig. S5). The deviation is mitigated as the decrease of the viscosity ratio, since imbibition dynamics reduces.

We have demonstrated unequivocally that wetting relative permeability significantly depends on imbibition dynamics. Now, the challenge lies in how to incorporate dynamics in the wetting relative permeability, and meanwhile relate the relative permeability to its quasi-static/steady-state counterpart which is measurable. In this work, we propose two models, and testify them against the results by the pore-network modeling. First, as usual, we fit the quasi-static relative permeability by the power-law model as $k_r^w = (s^e)^{1.1}$ as shown in Fig. S6 with the normalized saturation, $s^e = s/(1 - s_r)$, where $s_r$ is the residual saturation of 0.52 (notice that the wetting permeability is written as $k^w = 0.114(s^e)^{1.1}k_0$ where $k_0$ is the intrinsic permeability). A nearly linear relation between the wetting relative permeability and the normalized saturation is predicted for imbibition, which are also seen in some steady-state measurements (Akbarabadi & Piri, 2013; Ruprecht et al., 2014). In Fig. 6B, we observe a critical saturation value for each case of the viscosity ratio, below which the wetting phase does not form a conducting pathway. In light of this observation, we propose to scale the normalized saturation with respect to the critical saturation so that a final effective saturation is written as $s^{fe} = (s^e - s_c)/(1 - s_c)$. Then, we substitute the normalized saturation with the final effective saturation in the power-law model of quasi-static relative permeability, and set the relative permeability to zero whenever $s^{fe} < 0$. In such way, the wetting relative permeability for sintered glass beads under a high and a low viscosity ratios predicted by the power-law model is shown in Fig. 6C, which matches the relative permeability by the pore-network modeling very well. However, the critical saturation value should be pre-described, which obviously depends on imbibition dynamics. In order to fulfill the development of this permeability model, we need to further develop a relation between critical saturation and imbibition dynamics, which is beyond the scope of this paper.

Alternatively, motivated by the concept of non-equilibrium saturation in (Barenblatt et al., 2003) and the work of dynamic capillarity in (Zhuang et al., 2017), we propose the second model of wetting relative permeability as:

$$k_r^w = \left(s^e - m(s^e)^n \frac{ds^e}{dt}\right)^\alpha \tag{3}$$

where imbibition dynamics is represented by $ds^e/dt$, tau (i.e., τ) is assumed to be saturation-dependent as $m(s^e)^n$. We take two cases of the viscosity ratios of 55.9 and 0.1, to separately fit the two coefficients of $m$ and $n$. Moreover, we set the power exponent, α, to 1.1, in order to be consistent with the quasi-static permeability. Notice that we set the relative permeability to zero whenever $s^e - m(s^e)^n \frac{ds^e}{dt} < 0$ Fig. 6D shows the fitting results. Surprisingly, the case of viscosity ratio of 55.9 has the two coefficient values of m = 0.025 and n = −5.577, while the case of viscosity ratio of 0.1 has the two coefficient values of m = 0.022 and n = −5.714. This indicates that one set of coefficient values may cover a wide range of imbibition dynamics. Interestingly, the critical saturation can be well predicted by the proposed non-equilibrium permeability model, which is superior to the first proposed permeability model in this regard. A much longer media would be needed to cover a wide range of imbibition dynamics, which will be explored in the further study. Here, instead, we simply slow down imbibition dynamics by multiplying $ds^e/dt$ by a value smaller than unity, to mimic different strength of imbibition dynamics. Fig. S7 shows the effect of imbibition dynamics on the water relative permeability for sintered glass beads, in which air is the nonwetting phase. As expected, the water relative permeability approaches to its quasi-static curve as dynamics is weaker and weaker. Meanwhile, the critical water saturation reduces gradually. To sum up, we have seen that the proposed non-equilibrium permeability model (Eq. 3) can quantitatively describe the wetting relative permeability for the early stage of spontaneous imbibition. It also has the potential to be predictive for a wide/whole range of imbibition dynamics, which is crucial to the Darcy-scale modeling of spontaneous imbibition.

## 5. Conclusions and outlook

By using the efficient pore-network modeling, we have conducted a number of case studies of cocurrent spontaneous imbibition in sintered glass beads and Estaillades, under different contact angle values and viscosity ratios. The porous medium of sintered glass beads is homogenous, while Estaillades features strong pore-scale heterogeneity. In the case studies of spontaneous imbibition of water into the dry media, we found that pore-scale heterogeneity has significant influence on the entrapment of air, which in Estaillades is primarily due to the poor connectivity of pores, giving rise to the topological trapping. The REV size for spontaneous imbibition in Estaillades is much larger than that of a homogeneous medium. For both porous media imbibition rates can be adequately scaled with $\sqrt{cos\theta}$ ($\theta$ is the static contact angle) under the assumption of uniform wettability. The REV-scale capillary pressure locates in the tail of the quasi-static MIC (main imbibition curve).

We have shown that pore-filling events depend on imbibition dynamics and the viscosity ratio. In the case of quasi-static imbibition, the wetting process follows the mechanism of percolation-invasion. Snap-off and piston-type filling of pore throats give rise to severe entrapment of the nonwetting phase, which is even worse in heterogeneous Estaillades. In spontaneous imbibition, however, we observed the cofilling mechanism of small and large pores, which considerably reduces entrapment of the nonwetting. As the decrease of the viscosity ratio, we observed dry and partially-filled pores, with the majority of intermediate and large pores. This is due to the interplay of capillary force and viscous resistance of the nonwetting phase.

To upscale pore-scale imbibition dynamics, we have proposed two relative permeability models for the wetting phase in spontaneous imbibition. We have shown that both models can predict the permeability values by the pore-network modeling. The first model is to scale the normalized saturation with respect to the critical saturation as $k_r^w = ((s^e - s_c)/(1 - s_c))^\alpha$, where $s^e = s/(1 - s_r)$ is the normalized wetting saturation, $s_c$ is the critical saturation, and $\alpha$ is determined by the quasi-static relative permeability curve. Although this model is concise, the critical saturation value should be pre-described, which depends on imbibition dynamics. Alternatively, we developed the second model as $k_r^w = \left(s^e - m(s^e)^n \frac{ds^e}{dt}\right)^\alpha$, which is termed as the non-equilibrium model. The power exponent $\alpha$ is also determined by the quasi-static relative permeability curve. $m$ and $n$ are the fitting parameters. The developed non-equilibrium model can well predict the pore-network modeling results for the early stage of spontaneous imbibition under high and low viscosity ratios.

We need to further verify the model against lab experiments of core-scale spontaneous imbibition. Although direct measurements of relative permeability at different locations of a core sample are challenging, we can incorporate the non-equilibrium permeability model into the two-phase Darcy model, to verify if the two-phase Darcy model produces the same temporal saturation profiles and the broadening of wetting front as experimental data [ref]. To conduct the above verification, we further need to develop a non-equilibrium model for REV-scale capillary pressure. As discussion in Section 4.1 and shown in Fig. S3, imbibition dynamics does deviate capillary pressure away from its quasi-static values. Unlike relative permeability, capillary pressure may be directly measured in lab experiments.

Besides the verification by lab experiments, we may conduct core-scale (up to a few centimeters long at least) pore-network modeling of spontaneous imbibition. In such way, we can testify if the developed permeability model is applicable to a wide range of imbibition dynamics, such as the late stage of spontaneous imbibition. To this end, a parallelized pore-network simulator needs to be developed. Moreover, a hybrid imbibition model combining the pore-network modeling and the Darcy-scale modeling may be needed. Last but not least, the permeability model also needs to be verified for porous media of mixed-wettability, which are crucial to many practical applications.

**Acknowledgement**

C.Z.Q. acknowledges the support of National Natural Science Foundation of China (No. 12072053), and the support of Human Resources and Social Security Bureau of Chongqing (No. cx2020087).

# Supplementary Support

**Movie S1:** Spontaneous imbibition of water into the dry medium of sintered glass beads. The contact angle value is assumed to be 40°.

**Movie S2:** Spontaneous imbibition of water into the dry medium of Estaillades carbonate. The contact angle value is assumed to be 40°.

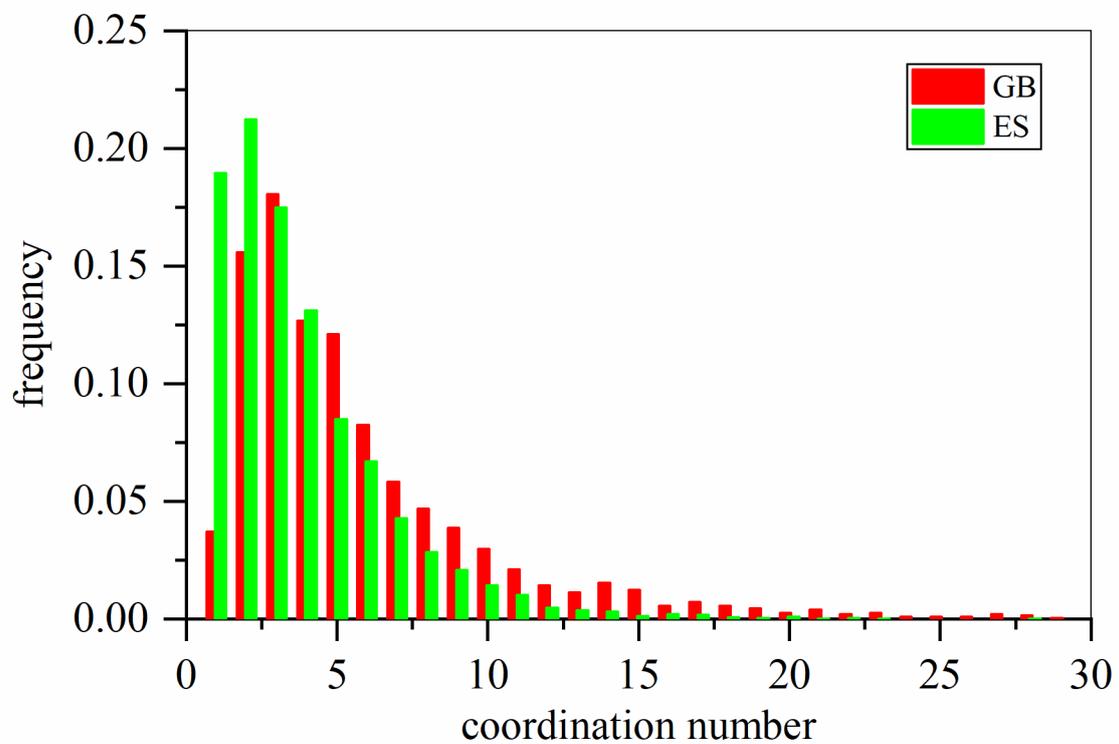

**Fig. S1:** Distributions of coordination number of sintered glass beads and Estaillades.

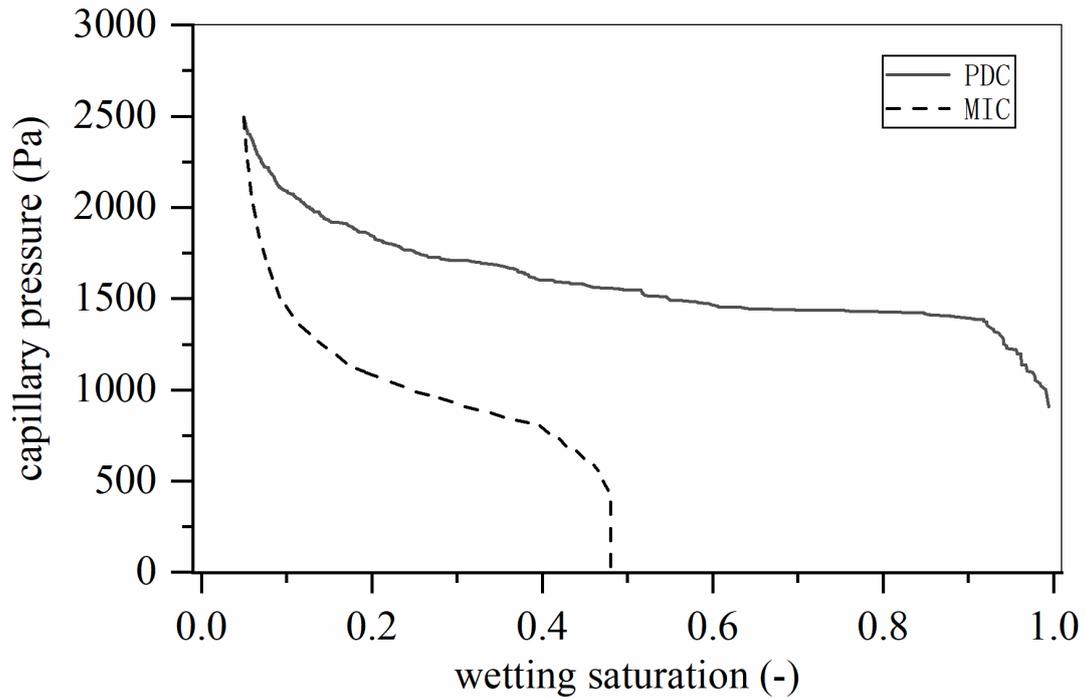

**Fig. S2:** Primary drainage curve (PDC) and main imbibition curve (MIC) of capillary pressure-saturation relationship for sintered glass beads. The static contact angle is 40°. As expected, the MIC is below the PDC due to the entrapment of nonwetting phase.

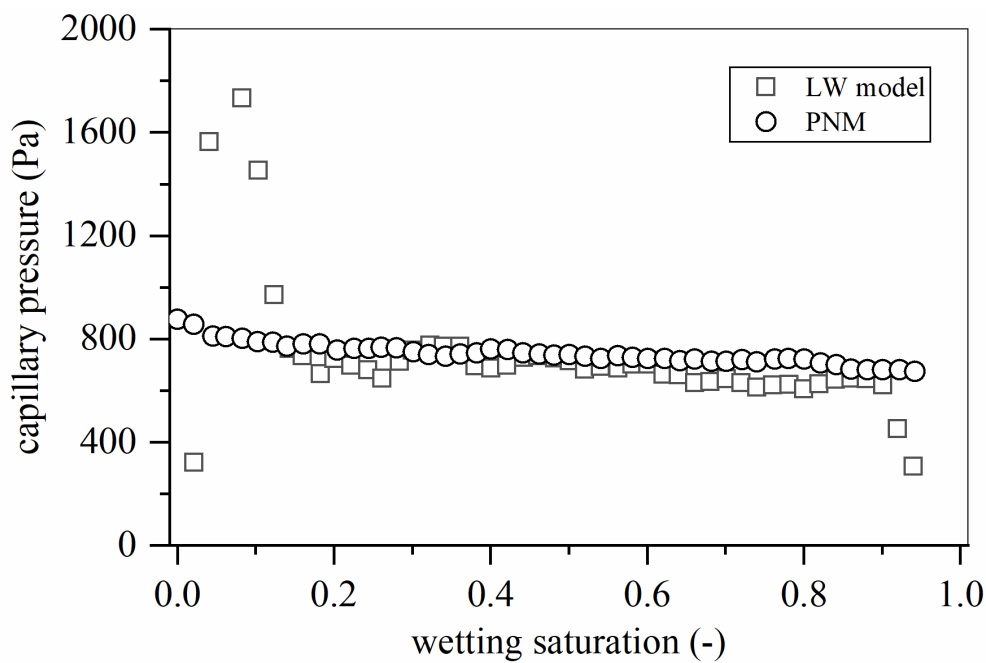

**Fig. S3:** Network-scale (or REV-scale) capillary pressure for sintered glass beads at different wetting saturation values fitted by the Lucas-Washburn (LW) model, and calculated by the pore-network model (PNM). The conventional LW model is modified to consider spontaneous imbibition in a dry porous medium.

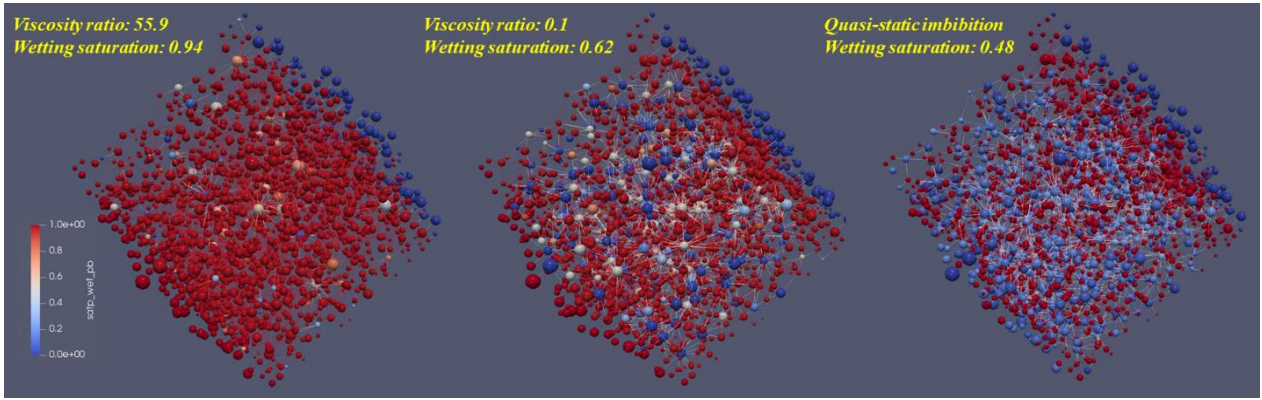

**Fig. S4:** wetting-phase distributions in sintered glass beads in the end of spontaneous imbibition and quasi-static imbibition.

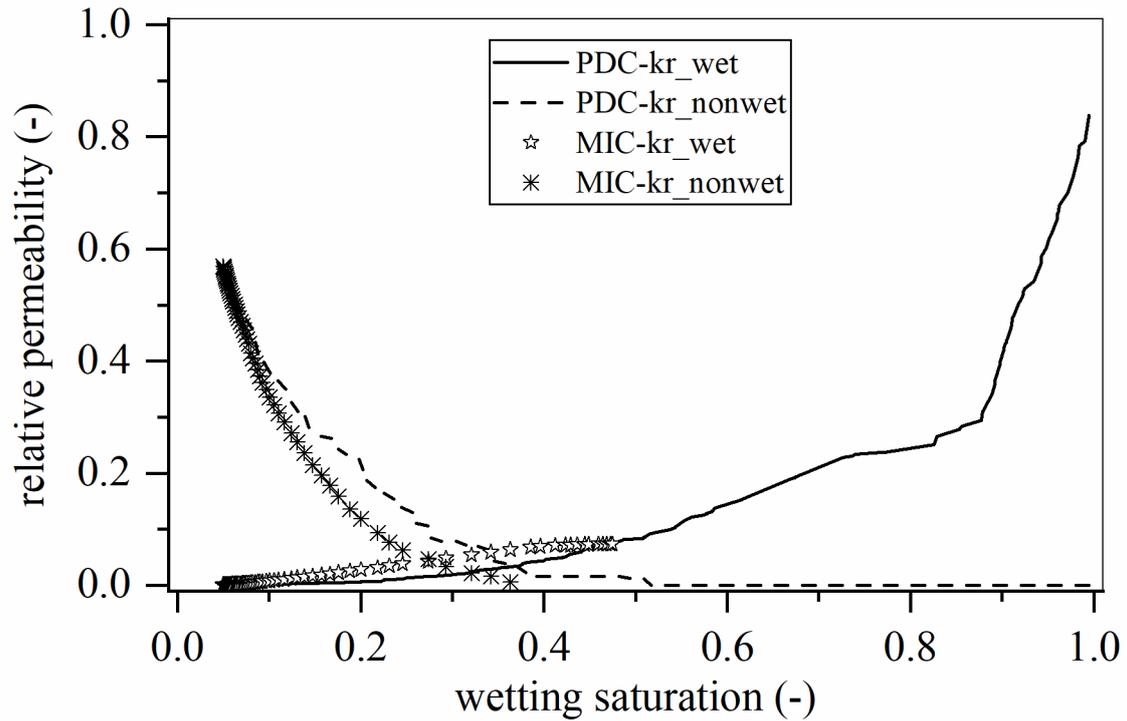

**Fig. S5:** Relative permeability of primary drainage and main imbibition for sintered glass beads by the PNM. For the nonwetting phase, the MIC (main imbibition curve) bellows the PDC (primary drainage curve), while for the wetting phase, the MIC is above the PDC. The contact angle is 40°.

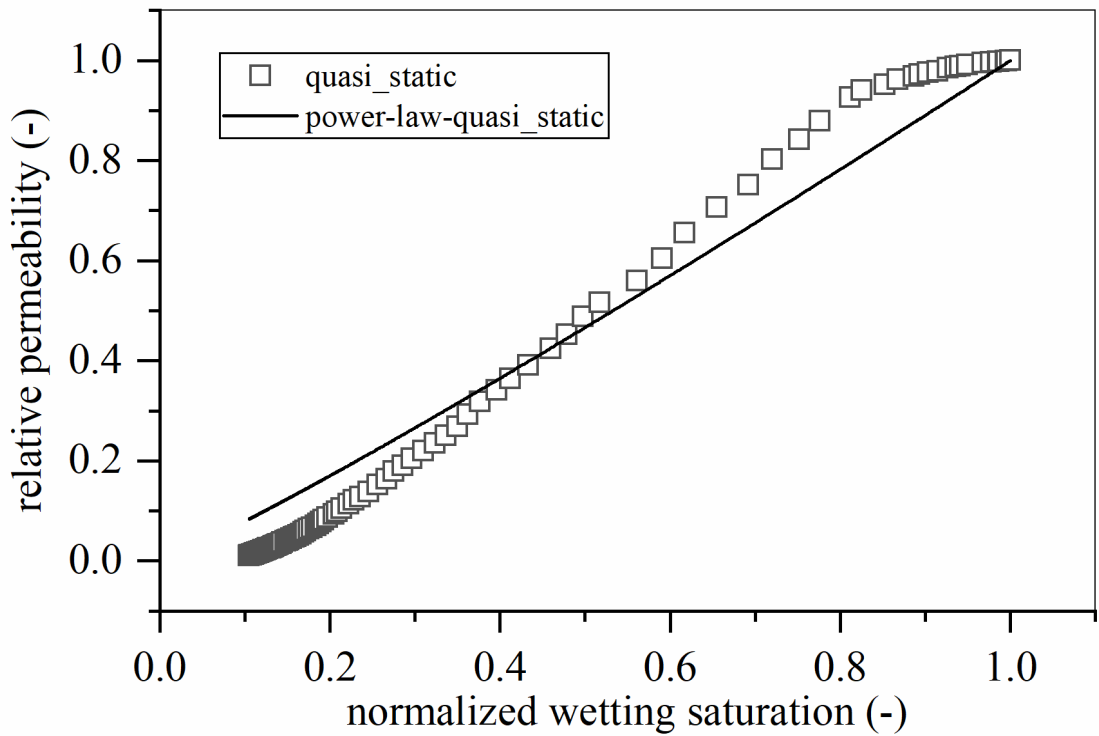

**Fig. S6:** Wetting relative permeability versus normalized wetting saturation (i.e., $s/(1-s_r)$) by the quasi-static PNM, and by the power-law model with the exponent of 1.1. The residual saturation, $s_r$, is 0.52. The contact angle is 40°.

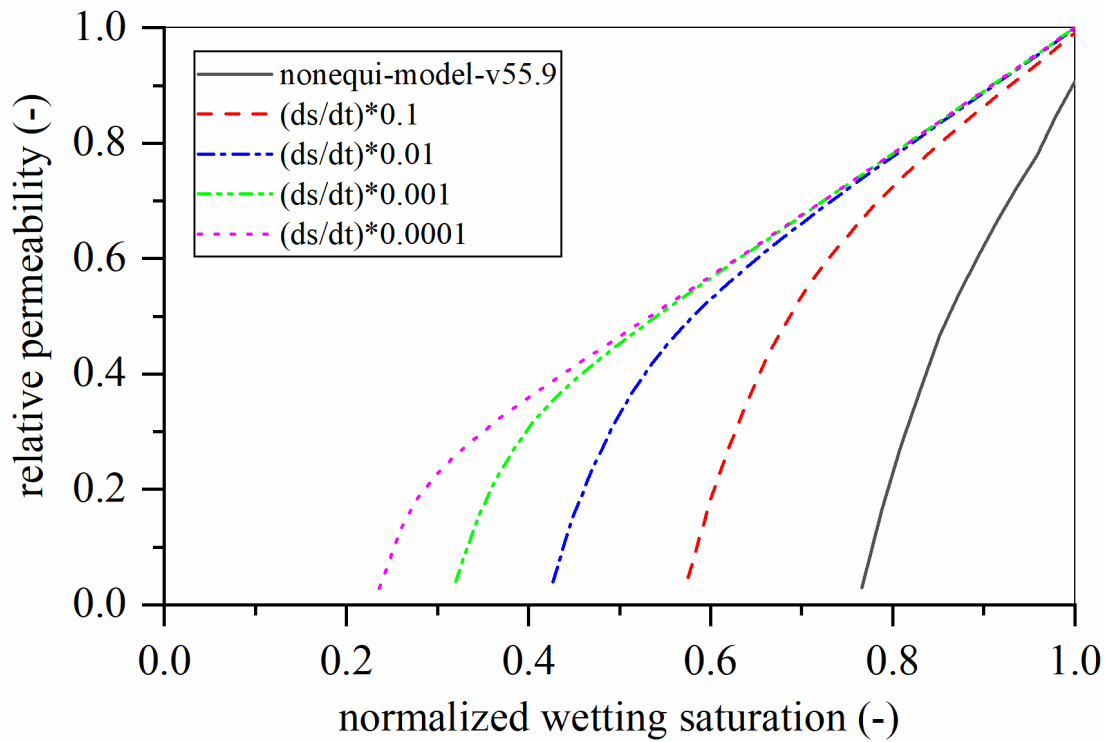

**Fig. S7:** Effect of imbibition dynamics on the water relative permeability for sintered glass beads. The used non-equilibrium model is $k_r^w = \left(s^e - 0.025(s^e)^{5.577}\frac{ds^e}{dt}\right)^{1.1}$, with the data of $ds^e/dt$ from the pore-network modeling.